\documentclass[%
 reprint,
superscriptaddress,
 amsmath,amssymb,
 aps,
]{revtex4-2}

\usepackage{graphicx}
\usepackage{dcolumn}
\usepackage{bm}
\usepackage{cellspace}

\begin{document}

\title{Collisionality scaling of the tokamak density limit}

\author{A. D. Maris}
 \email{andrew.maris@columbia.edu}
\affiliation{Massachusetts Institute of Technology (MIT), Plasma Science and Fusion Center (PSFC)}
\affiliation{Columbia University, Department of Applied Physics and Applied Mathematics}
\author{C. Rea}
\affiliation{Massachusetts Institute of Technology (MIT), Plasma Science and Fusion Center (PSFC)}
\author{A. Pau}
\affiliation{École Polytechnique Fédérale de Lausanne (EPFL), Swiss Plasma Center (SPC)}
\author{R. Granetz}
\affiliation{Massachusetts Institute of Technology (MIT), Plasma Science and Fusion Center (PSFC)}
\author{E. Marmar}
\affiliation{Massachusetts Institute of Technology (MIT), Plasma Science and Fusion Center (PSFC)}
\author{the Alcator C-Mod team}
\altaffiliation[See author list of \textit{E. S. Marmar et al., 2015 Nucl. Fusion 55 104020 https://doi.org/10.1088/0029-5515/55/10/104020}]{}\noaffiliation
\author{the ASDEX Upgrade team}
\altaffiliation[See author list of \textit{T. Pütterich et al., 2026 Nucl. Fusion 66 116002 https://doi.org/10.1088/1741-4326/ae61c8}]{}\noaffiliation
\author{the DIII-D team}
\altaffiliation[See author list of \textit{C. T. Holcomb et al., 2024 Nucl. Fusion 64 112003 https://doi.org/10.1088/1741-4326/ad2fe9}]{}\noaffiliation
\author{JET Contributors}
\altaffiliation[See author list of \textit{C. F. Maggi et al., 2024 Nucl. Fusion 64 112012 https://doi.org/10.1088/1741-4326/ad3e16}]{}\noaffiliation
\author{the TCV team}
\altaffiliation[See author list of \textit{C. Theiler et al., 2026 Nucl. Fusion 66 116007 https://doi.org/10.1088/1741-4326/ae6d12}]{}\noaffiliation
\author{the EUROfusion Tokamak Exploitation team}
\altaffiliation[See author list of \textit{N. Vianello et al., 2026 Nucl. Fusion 66 116010 https://doi.org/10.1088/1741-4326/ae71ec}]{}\noaffiliation

\date{\today}

\begin{abstract}
The threshold at which density-limiting plasma instabilities occur in tokamak fusion reactors is not well understood. Here, we present the first large-scale empirical evidence of a common collisionality threshold present across multiple devices and confinement regimes. Machine-learning-guided analysis of a diverse experimental database identifies a scaling in edge effective collisionality ($\nu_*$) that bears resemblance to the widely utilized Greenwald limit but yields a substantially lower false positive rate. This threshold decreases with dimensionless pressure ($\beta$), consistent with theoretical expectations for enhanced transport from electromagnetic effects on resistive ballooning mode turbulence. These results suggest that future burning plasma devices such as ITER and ARC could potentially achieve higher density, and thus fusion power, than previously expected.
\end{abstract}

\maketitle

Plasma density is a key parameter in nuclear fusion, as fusion power scales with the square of the density ($P_{\rm fus} \propto n^2$). In tokamak reactors, the maximum achievable density is limited by an anomalous collapse of the plasma edge temperature, which results in either a major loss of energy confinement or a magnetohydrodynamic (MHD) instability. The physical mechanism responsible for this ``density limit'' remains contested. In practice, the operational boundary is often estimated using the empirical Greenwald limit~\cite{greenwald1988new},
\begin{equation}
n_G = \frac{I_p }{\pi a^2},
\end{equation}
where $n_G$ is the predicted maximum line-averaged density (in units of $10^{20}\,\mathrm{m}^{-3}$), $I_p$ is the plasma current (units of MA), and $a$ is the minor radius (units of m).
The Greenwald limit was originally identified as an approximate, empirical upper bound on the operational range of tokamak density, rather than as a precise instability predictor.
For this reason, the scaling captures general trends across devices, but deviations of 50\% or more are observed in some circumstances \cite{bell1992attainment,stabler1992density,pucella2013density,de1998influence,osborne2001gas,greenwald2002density,maingi2005review,huber2013impact, lang2014elm,pucella2017analytical,ding2024high,sauter2025operation,maris2025correlation}.
This leads to uncertainty regarding the expected performance of future experiments, such as ITER \cite{ikeda2007progress} and ARC \cite{hillesheim2026overview}, and requirements for their mission-critical instability avoidance schemes \cite{maraschek2017path,sieglin2023disruption}.
Here, we take a complementary approach: we explicitly optimize a scaling to discriminate density limit precursor states from stable operation.

Experimental \cite{greenwald1988new,labombard2001particle}, numerical \cite{rogers1998phase,xu2003transitions}, and theoretical studies \cite{eich2021separatrix,singh2022zonal,giacomin2022first} have suggested that transport processes play a central role in setting the density limit. However, past experimental validations are limited in scope, often focusing on individual devices or confinement regimes, and lack quantitative assessment of the false positive rate. Additionally, correlations with radiative phenomena such as MARFEs \cite{lipschultz1984marfe} and X-point radiators \cite{bernert2021x} make it difficult to isolate the underlying mechanism.

\begin{figure}
    \centering
    \includegraphics[width=\linewidth]{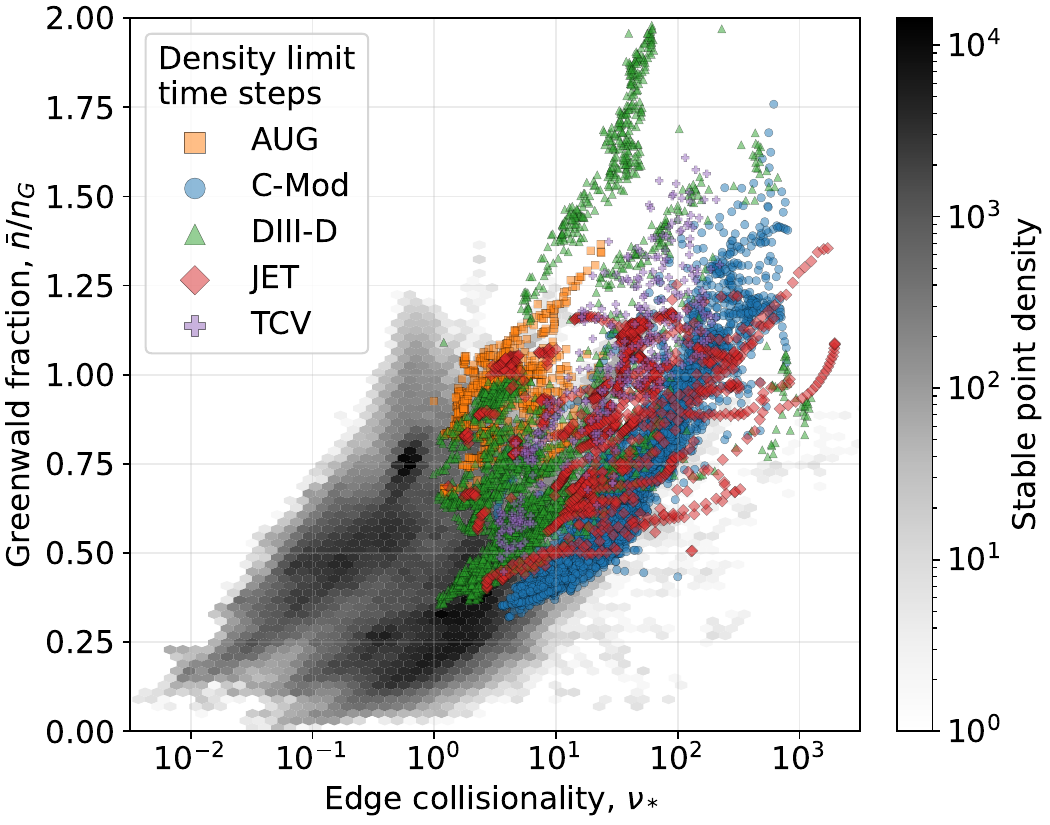}
    \caption{Data points associated with the precursor phase of the density limit for five tokamaks plotted in terms of edge collisionality ($\nu_*$) and Greenwald fraction. Stable data points are represented by gray hexbins. Nearly all density limit time points lie in the collisional regime, $\nu_* > 1$.}
    \label{fig:coll_gwf_hex}
\end{figure}

In this Letter, we provide the first large-scale, multi-device experimental evidence of how edge collisionality plays a central role in setting the tokamak density limit across confinement regimes. We analyze a newly assembled database spanning several tokamaks: Alcator C-Mod, ASDEX Upgrade (AUG), DIII-D, JET, and TCV (illustrated in Fig.~\ref{fig:coll_gwf_hex}). These experiments span a wide range in density ($\bar{n}_e = 0.1$--$2.7\cdot 10^{20}$~m$^{-3}$), magnetic field ($B_T = 1.4$--$7.6$~T), device size ($R_0 = 0.68$--$2.9$~m), plasma heating sources (ohmic, NBI, ICRF), triangularity (positive and negative), and wall material (carbon, molybdenum, tungsten, beryllium). There are 82 density limits in H-mode, 243 in L-mode, and 4,359 stable discharges gathered via the DEFUSE \cite{pau2023modern} and DisruptionPy \cite{trevisan2026disruptionpy} frameworks. As with previous analyses \cite{maris2025correlation,maris2026real}, the precursor phase for L-mode discharges is defined as the points in time between the formation of the MARFE and the point at which the MARFE is destabilized. For H-mode density limits, the precursor phase is determined by a fixed time window before the H/L back-transition (100 ms for the larger devices---DIII-D, AUG, JET---and 30 ms for smaller devices with faster inherent time scales---TCV, C-Mod). We have excluded discharges where there are disruptions or macroscopic instabilities caused by other mechanisms or that cannot be clearly associated with a ``pure'' density limit. We also ignore discharge phases where diagnostic coverage is limited, including the dynamic ramp-up and ramp-down stages at the very beginning and end of the experiments.

We apply an interpretable machine learning workflow to identify power laws that accurately separate the stable and density limit plasma states. The algorithm searches over a standard set of dimensionless variables chosen to parameterize edge turbulence: edge effective collisionality \(\nu_* = \nu_{ei} q_{*} R_0/ (v_{\rm th} \epsilon^{3/2}) \), edge dimensionless pressure \(\beta = p_{\rm edge} / (B^2/2\mu_0)\), edge safety factor \(q_{\rm *} = \frac{2 \pi}{\mu_0} \frac{B_T a^2}{I_p R_0} \frac{1 + \kappa^2 }{2}\), and normalized edge poloidal gyroradius \(\rho_{*,\rm pol} = m_D v_{\rm th}/(eB_{\rm pol} a)\). We crudely assume a purely deuterium plasma as edge $Z_{\rm eff}$ measurements are not available across the database. We also include two other dimensionless parameters related to radiation: normalized edge ionization energy \(T_*= T_{\rm edge}/E_{\rm ion}\) and global radiated fraction \(f_{\rm rad} =P_{\rm rad}/P_{\rm in}\). Edge quantities are evaluated using Thomson scattering measurements averaged over the radial interval $0.85 \le \rho_{\mathrm{pol}} < 0.95$, as in previous analyses \cite{giacomin2022first,maris2025correlation}.

\begin{figure}
    \centering
    \includegraphics[width=\linewidth]{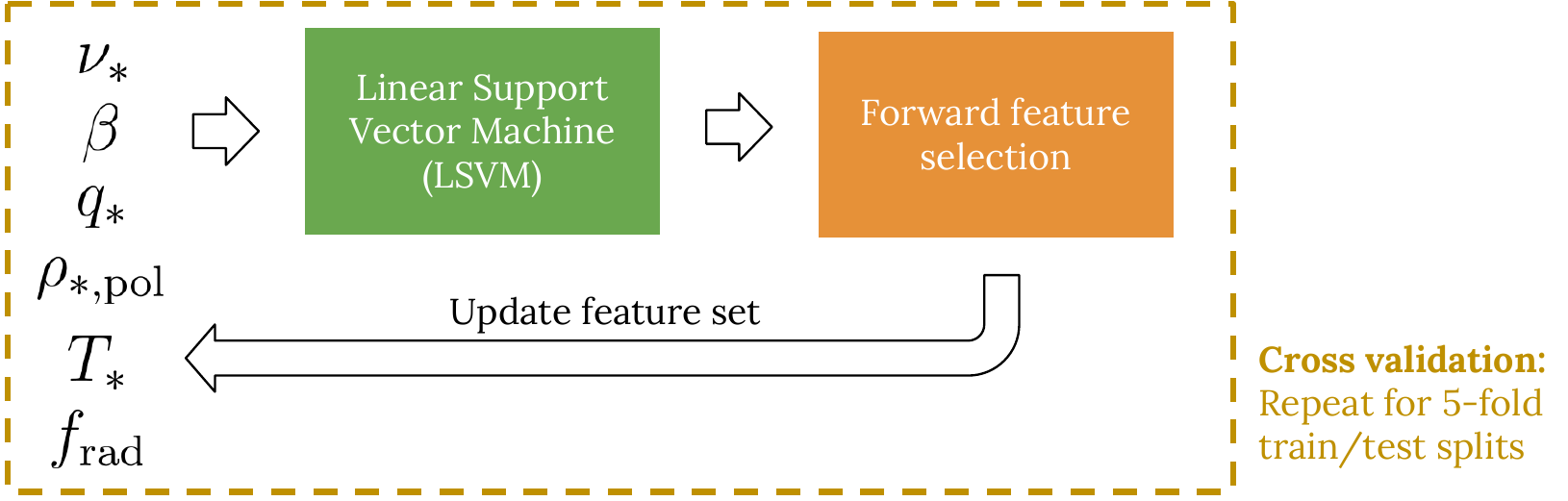}
    \caption{Candidate power-law solutions are explored via Linear Support Vector Machine (LSVM) classification and forward feature selection. Results are reported over five-fold train/test splits.}
    \label{fig:MLWorkflow}
\end{figure}

The machine learning workflow used in this study is summarized in Fig.~\ref{fig:MLWorkflow}. For a given set of variables, a Linear Support Vector Machine (LSVM) fits a linear classification boundary to the data. Notably, LSVMs have previously been used to identify a physically interpretable boundary for the computational fidelity of a class of simulations \cite{murillo2020machine}; here, we apply this algorithm to find an accurate power law decision boundary. Model selection is performed using forward feature selection, and five-fold cross-validation is applied to provide error bars on predictive output. The primary metric for assessing model performance is the area under the receiver operating characteristic curve (AUC), a standard metric that quantifies classification performance across sensitivity levels. Predictions are evaluated on a per-discharge basis, as is standard in instability prediction studies, rather than on a per-time-point basis.

As a baseline, we evaluate the predictive capability of the Greenwald fraction.
We find that the line-averaged Greenwald fraction achieves an AUC of $0.918\pm0.012$, which is a moderate degree of prediction accuracy.
We further find that the edge density variant of the Greenwald fraction is a less accurate descriptor of the data, with an AUC of $0.853\pm0.015$.
We do not find an improvement in prediction accuracy from re-optimizing the Greenwald fraction variables ($\bar{n}, I_p, a$) by training an LSVM on the dataset.
Fitting a heating power dependence likewise yields no significant improvement (LSVM-optimized scaling: $\bar{n}^{\rm limit} \sim n_G P_{\rm in}^{0.19}$, AUC = $0.922 \pm 0.011$).

\begin{figure}
    \centering
    \includegraphics[width=\linewidth]{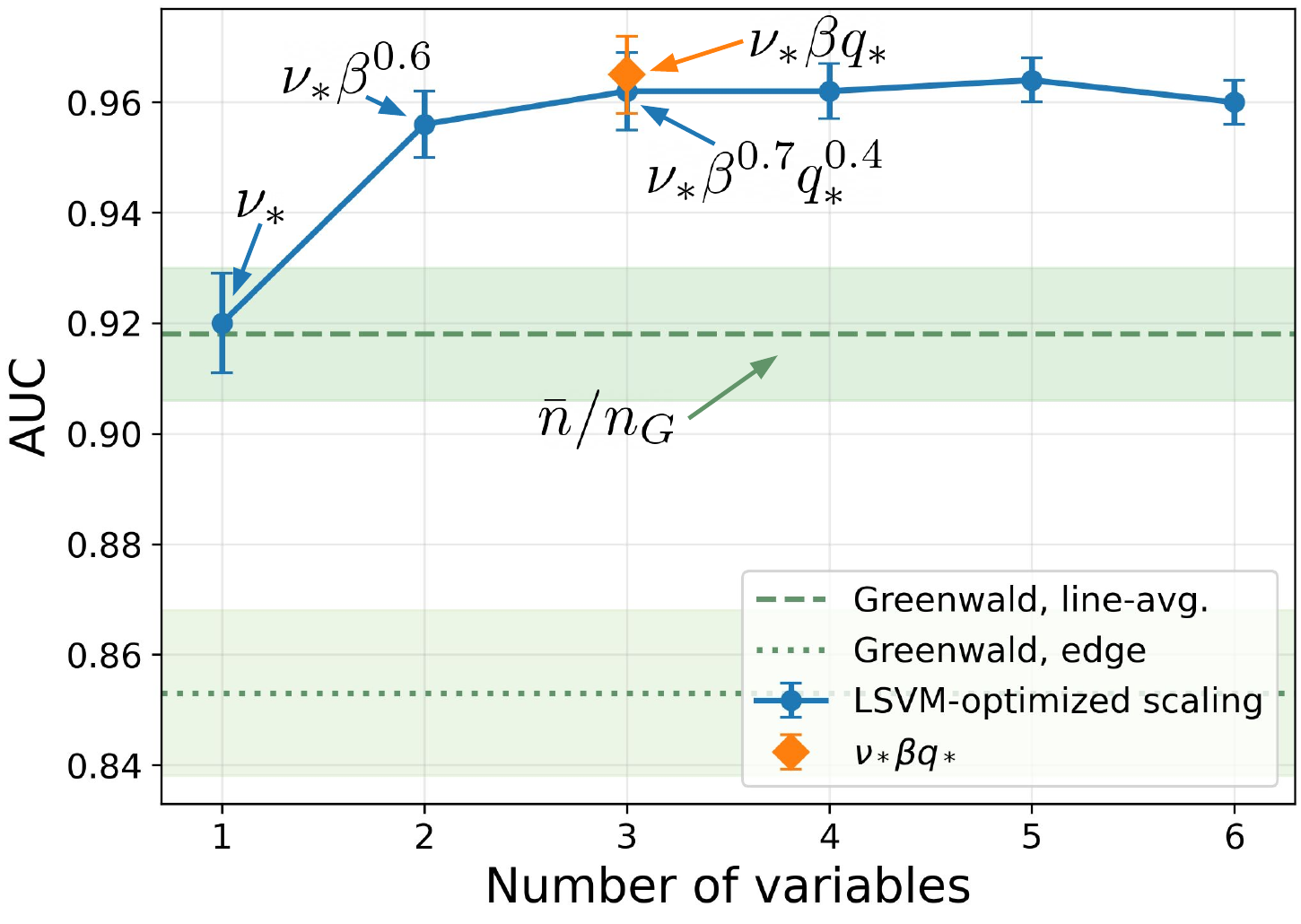}
    \caption{Predictive accuracy of candidate density limit scalings, including the line-averaged Greenwald fraction (green, dashed), edge Greenwald fraction (green, dotted), scalings from the LSVM (blue), and the $\nu_* \beta q_*$ scaling (orange). The largest improvement occurs when adding collisionality and normalized pressure, and performance saturates after three parameters. The simple physics-motivated metric $\nu_* \beta q_*$ achieves comparable accuracy to the ML-optimized scalings.}
    \label{fig:feature_selection}
\end{figure}

\begin{figure}
    \centering
    \includegraphics[width=\linewidth]{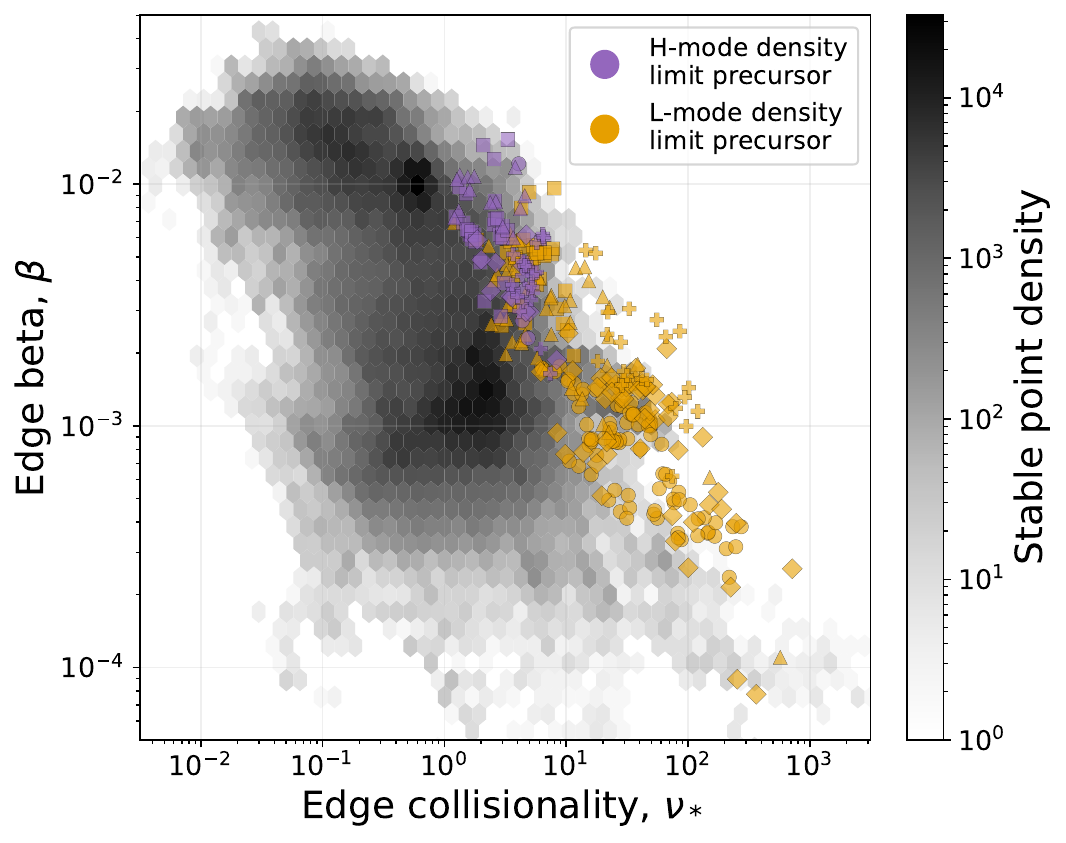}
    \caption{The edge collisionality-pressure operating space of this dataset. Stable data time points are plotted in gray hexbins, with density limit precursor time points plotted on top (averaged per shot for visualization, marker shape same as in Fig.~\ref{fig:coll_gwf_hex}). The density limit points cluster at the high-collisionality, high-$\beta$ edge of the operating space.}
    \label{fig:coll_beta}
\end{figure}

The results of the LSVM-guided search are summarized in Fig.~\ref{fig:feature_selection}. The first parameter selected by the model is the edge effective collisionality, $\nu_*$. Remarkably, this single parameter achieves predictive accuracy comparable to the line-averaged Greenwald fraction. The preference for collisionality is insensitive to the choice of input variables; including edge density as an independent variable does not improve prediction accuracy. Figure~\ref{fig:coll_gwf_hex} illustrates that the density limit occurs in the collisional regime, as nearly all precursor data points satisfy $\nu_* > 1$. Following collisionality, the second parameter selected is $\beta$, which significantly improves predictive accuracy to an AUC of $0.956 \pm 0.006$.
From the state space of these parameters shown in Fig.~\ref{fig:coll_beta}, we make two observations: first, that the density limit occurs at the high-collisionality edge of the operating space, and second, that the collisionality threshold at which the density limit occurs decreases with $\beta$.
We also note that the L-mode and H-mode density limit events form a single continuous band in this state space.
The third parameter selected is the edge safety factor, leading to an AUC of $0.962 \pm 0.007$.
As shown in Fig.~\ref{fig:feature_selection}, additional parameters beyond this point provide no improvements in predictive accuracy, including the radiation-related variables ($T_*$ and $f_{\mathrm{rad}}$).

The central role of collisionality in setting the density limit is consistent with expectations from fluid models of the plasma edge and past experimental observations. Prior numerical analysis identified the onset of catastrophic transport in high-collisionality and high-$\beta$ regimes, hypothesized to be the origin of the density limit \cite{rogers1998phase}. Later experimental studies found the density limit in both H- and L-mode correlated with enhanced transport at high edge collisionality \cite{suttrop1997identification,labombard2001particle,sun2021role,wang2026evolution,miller2026experimental}. Low edge temperature \cite{petrie1993plasma,rapp1999density,yamada2007density} and low electron adiabaticity \cite{hong2017edge,hong2026disentangling} are also reported when approaching the density limit, consistent with a collisionality-based mechanism.

Recent theoretical work proposed that strongly enhanced energy transport should occur when the characteristic wavenumber of resistive ballooning modes $k_{\rm RBM} = \sqrt{\alpha_{c}/ 2 C \sqrt{\omega_B}} \sim (\nu_* q_*)^{-1/2}$ at the separatrix matches the electromagnetic wavenumber $k_{\rm EM} = \sqrt{\beta_e/\mu} \sim \beta^{1/2}$ \cite{eich2021separatrix}, where $C \equiv 0.51 \mu \nu_{ei} (q_*R)^2/(c_s \lambda_\perp)$ is a normalized collision frequency and $\omega_B \equiv 2\lambda_\perp/R$ is a normalized perpendicular gradient length. This mechanism has been proposed to explain the density limit in L-mode. This critical threshold can be expressed as
\begin{equation} \label{eq:SepOSThres}
    \nu_*\beta q_*\big|_{\rm sep}  > \frac{1.4 \alpha_c \mu^{1/2}}{\epsilon^{3/2}} \sqrt{\frac{\lambda_\perp}{R_0}} \approx 0.3 \sqrt{\frac{\lambda_\perp}{R_0}},
\end{equation}
where the right-hand side is approximated under typical conditions ($\epsilon \approx 1/3$, $\alpha_c \approx 2.4$ \cite{eich2021separatrix}).  The left-hand side can also be written in terms of the turbulence parameter $\alpha_t \equiv (1 + \tau_i)C\omega_B \sim \nu_* q_*$ \cite{eich2021separatrix} of the drift-Alfv\'{e}n (DALF) model. Beyond this critical threshold, electromagnetic effects on resistive ballooning modes can increase transport both directly, through magnetic flutter, and indirectly, as induction weakens the diamagnetic stabilization provided by the parallel electron response \cite{manz2025turbulence}.

Motivated by this prediction, we test the simplified scaling \(\nu_* \beta q_*\) on this dataset. This yields an AUC of $0.965 \pm 0.007$, which is as accurate as the best LSVM-optimized scalings and a significant improvement upon the Greenwald fraction.
When calibrated to a true positive rate of 95\% (the ITER operational requirement for thermal quench prediction \cite{de2016requirements}), the false positive rate of \(\nu_* \beta q_*\) is 1.6$\times$ lower than the Greenwald fraction (16.6\%, compared with 27.0\%); at a true positive rate of 90\%, this improvement increases to 2.6$\times$ (8.6\%, compared with 22.2\%).
Applied to the L-mode and H-mode density limit data separately, \(\nu_* \beta q_*\) is likewise a better predictor than the Greenwald fraction and on par with the best LSVM-optimized scaling for each regime.

\begin{figure}
    \centering
    \includegraphics[width=0.85\linewidth]{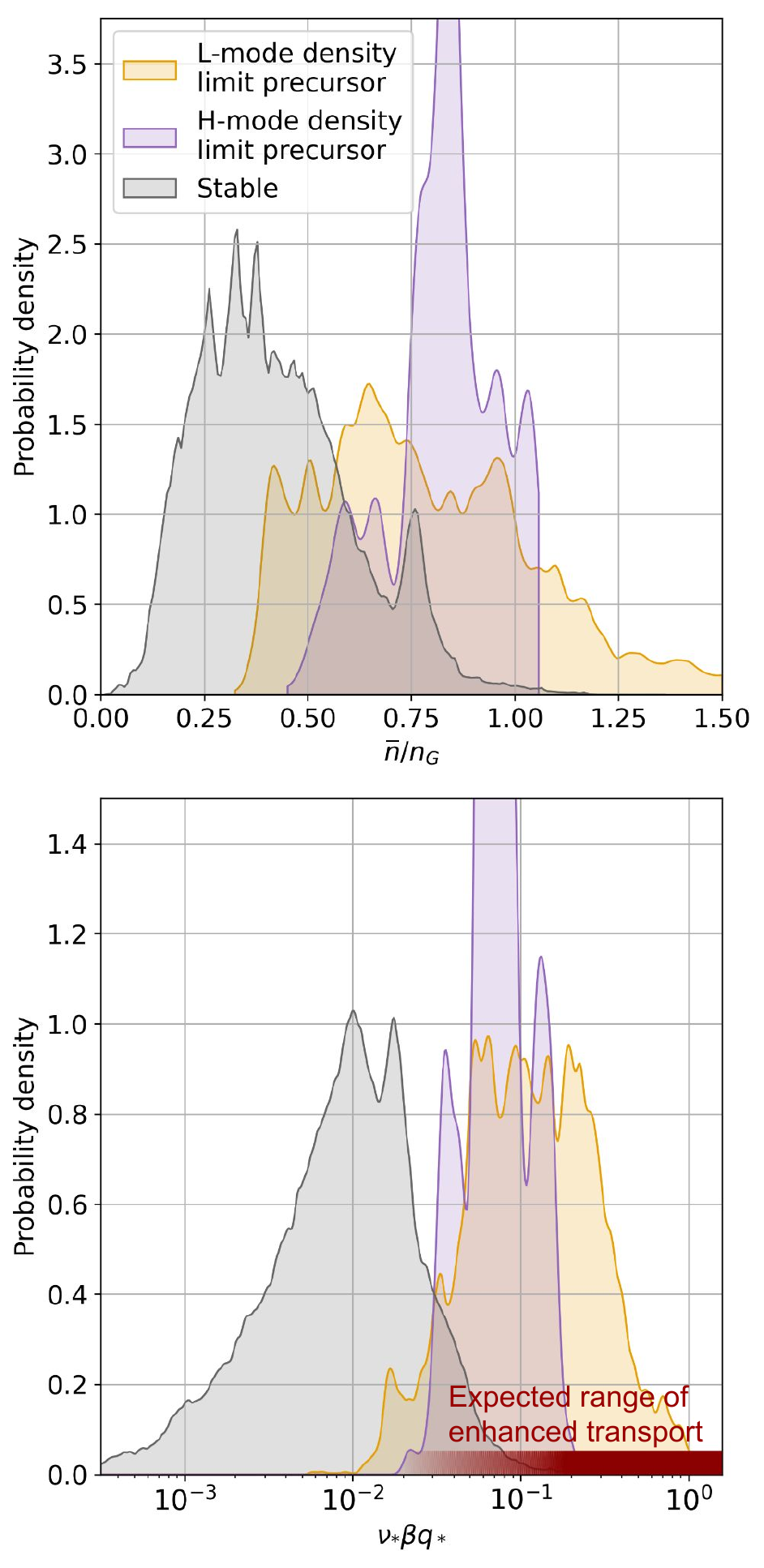}
    \caption{Distribution of stable, L-mode density limit precursor, and H-mode density limit precursor data points plotted in both Greenwald fraction and $\nu_* \beta q_*$. The greater predictive power of $\nu_* \beta q_*$ is reflected in the smaller overlap between stable and density limit points. Also shown is the expected regime (red) where enhanced transport should begin under typical conditions approaching the density limit: $n_{\rm edge} \approx 2 \cdot n_{\rm sep}$ \cite{manz2023power}, $\alpha_c \approx 2.4$ \cite{eich2021separatrix}, and $\lambda_\perp /R_0 \approx 3 \cdot 10^{-4}$--$3 \cdot 10^{-2}$.}
    \label{fig:KDE}
\end{figure}

Figure~\ref{fig:KDE} visually compares how \(\nu_* \beta q_*\) and the Greenwald fraction separate the density limit and stable data in this study.
The Greenwald fraction provides a moderate degree of separation between the stable and density limit regimes; this is to be expected as this parameter was derived as a rule-of-thumb rather than a precise instability predictor. The physically motivated \(\nu_* \beta q_*\) scaling produces a substantially clearer distinction between the stable and unstable regimes.
We also plot the expected range of enhanced transport under typical conditions approaching the density limit: $n_{\rm edge} \approx 2 \cdot n_{\rm sep}$ \cite{manz2023power}, $\alpha_c \approx 2.4$ \cite{eich2021separatrix}, and a conservative range of $\lambda_\perp /R_0 \approx 3 \cdot 10^{-4}$--$3 \cdot 10^{-2}$.
There is agreement between the expected range where enhanced transport should occur, \(\nu_* \beta q_* |_{\rm edge} \gtrsim 0.02\)--\(0.2 \), and the observed range of density limit precursor states.

The \(\nu_* \beta q_*\) threshold is compatible with decades of experimental observations of a density limit that generally scales like the Greenwald limit. Rewriting in terms of edge density, a modified Greenwald scaling emerges,
\begin{equation} \label{eq:modGreenwald}
\nu_* \beta q_* \sim
\left(
\frac{n}{n_G}
\frac{1}{T^{1/2}}\frac{1+\kappa^2}{ a^{1/2} \epsilon^{1/4}}
\right)^2,
\end{equation}
where $n$ is the edge density, $\kappa$ is the plasma elongation, $T$ is the edge temperature, and $\epsilon$ is the inverse aspect ratio.
In this expression, the edge Greenwald fraction $n/n_G$ is a key parameter, with additional dependencies on temperature, machine size, and shaping. The edge temperature dependence ($n^{\rm limit} \sim T^{1/2}$) is the most consequential deviation from the standard Greenwald limit, as this leads to a density limit that grows with edge temperature and hence, in general, with heating power.
In the parallel conduction-limited regime ($T_{\rm sep}\sim P_{\rm SOL}^{2/7}$), the threshold on edge density scales like $P_{\rm SOL}^{1/7}$. More generally, the response of the edge temperature to total heating power can vary between plasma scenarios, which may explain why a range of heating power sensitivities has been reported \cite{duesing1986first,stabler1992density,bell1992attainment,petrie1993plasma,de1998influence,huber2013impact,bernert2014h,huber2017comparative,schmitz2023accessing,sauter2025operation,maris2026real} and why no single heating power scaling improves the prediction accuracy of the Greenwald fraction across the database.
The additional shaping quantities vary modestly across operating scenarios, leading to little effect on the density limit threshold.

We note two deviations from the original theoretical predictions. First, we evaluate the quantity \(\nu_* \beta q_*\) utilizing density and temperature measurements in the region $0.85 \leq \rho_{\rm pol} < 0.95$, as opposed to at the separatrix as prescribed by theory. However, it has been shown that the edge and separatrix conditions approach each other near the density limit, even in H-mode \cite{labombard2001particle,bernert2014h,manz2023power}. Additionally, this definition significantly eases data collection, enabling the large statistics of this study and opening the door to real-time plasma control schemes \cite{maris2026real}. The second departure from the original theory is that this scaling unexpectedly predicts the H-mode density limit as well.
While theoretical treatments of the density limit generally prescribe mechanisms specific to L-mode or H-mode conditions \cite{zanca2019power,eich2021separatrix,singh2021bounds,brown2021generalization,stroth2022model,giacomin2022first}, past experimental observations of a Greenwald-limit-like threshold in both regimes and the shared threshold in $\nu_* \beta q_*$ observed in this study are consistent with a common mechanism causing enhanced transport.

Projecting to future tokamaks, these results suggest robust stability to the density limit in burning plasmas.
When organized in terms of the Greenwald fraction, the estimated operating point of ITER ($\bar{n}/n_G = 0.85$) is well within the range of observed H-mode and L-mode density limit cases in present experiments.
In contrast, the high edge temperature of ITER's intended operating regime ($n_{\rm edge} \approx 8\cdot10^{19}\,\mathrm{m}^{-3}$ and $T_{\rm edge} \approx 6$ keV \cite{koechl2020evaluation}) results in relatively low values of $\nu_* \beta q_*$, far from where density limit events occur in the database.
This is true for several major tokamak reactor concepts examined, including ARC \cite{hillesheim2026overview}, MANTA \cite{rutherford2024manta}, and EU-DEMO \cite{federici2019overview}; all have relatively high Greenwald fractions ($\bar{n}/n_G > 0.85$) but low edge $\nu_* \beta q_*$ values ($\nu_* \beta q_* \ll 10^{-2}$), suggesting a substantial margin of stability to the density limit.
While extrapolation to reactor regimes must be treated with caution, this suggests that higher densities, and therefore fusion power, may be possible.

Further analysis can enhance our understanding of this important operational limit.
Although the $\nu_* \beta q_*$ metric is an improved predictor of the density limit across the multi-machine dataset studied here, some misclassification between stable and density limit states remains, particularly in H-mode.
This could be a result of other mechanisms \cite{manz2023power,manz2025density} playing a secondary role or due to sensitivities in the theoretical model that are not captured in the present analysis, such as isotope mass, gradient scale length, and ion dilution.
Moreover, the edge conditions used here represent averages over a region, not precise measurements at the separatrix.
While this definition facilitates a consistent multi-device comparison, as the location of the separatrix can be challenging to determine, future studies can probe this sensitivity.

In summary, we find that the onset of the tokamak density limit across a multi-device (Alcator C-Mod, AUG, DIII-D, JET, and TCV) experimental database is well described by a combination of three dimensionless parameters measured in the plasma edge:\;$\nu_*$,\;$\beta$,\;and\;$q_*$.
A machine-learning-guided search of the dataset selects these three parameters to maximize the separation between stable and density limit precursor states.
A simple scaling in these parameters, \(\nu_* \beta q_*\), previously predicted to correspond to a threshold for enhanced transport \cite{eich2021separatrix}, achieves higher predictive power than the Greenwald fraction in terms of AUC ($0.965\pm0.007$ vs. $0.918\pm0.012$) and false positive rate (1.6$\times$ lower at a true positive rate of 95\% and 2.6$\times$ lower at 90\%).
The $\nu_* \beta q_*$ scaling is consistent with past observations of a Greenwald-like density limit, with a heating power sensitivity entering through the edge temperature.
This work illustrates that precise measurements of the separatrix are not needed for improved density limit prediction, substantially easing requirements for real-time instability avoidance.
The agreement between the empirical trends and theoretical predictions provides new experimental support for models in which the density limit arises from enhanced turbulent transport.
These results suggest that the density limit may be more fundamentally understood as a threshold in edge collisionality and $\beta$, rather than simply a critical density, with important implications for the achievable performance of future burning plasma devices.

\begin{acknowledgments}
The authors would like to thank D. Humphreys, A. Hubbard, J. Hughes, and N. Logan for providing insightful feedback on this study. We also thank J. Dunsmore and A. Miller for guidance in gathering C-Mod Thomson scattering data.  This material is based upon work supported by the U.S. Department of Energy, Office of Science, Office of Fusion Energy Sciences, using the DIII-D National Fusion Facility, a DOE Office of Science user facility, under Awards DE-FC02-04ER54698 and DE-SC0014264.

This work has been carried out within the framework of the EUROfusion Consortium, via the Euratom Research and Training Programme (Grant Agreement No 101052200 — EUROfusion) and funded by the Swiss State Secretariat for Education, Research, and Innovation (SERI). Views and opinions expressed are however those of the author(s) only and do not necessarily reflect those of the European Union, the European Commission, or SERI. Neither the European Union nor the European Commission nor SERI can be held responsible for them.

Disclaimer: This report was prepared as an account of work sponsored by an agency of the United States Government. Neither the United States Government nor any agency thereof, nor any of their employees, makes any warranty, express or implied, or assumes any legal liability or responsibility for the accuracy, completeness, or usefulness of any information, apparatus, product, or process disclosed, or represents that its use would not infringe privately owned rights. Reference herein to any specific commercial product, process, or service by trade name, trademark, manufacturer, or otherwise does not necessarily constitute or imply its endorsement, recommendation, or favoring by the United States Government or any agency thereof. The views and opinions of authors expressed herein do not necessarily state or reflect those of the United States Government or any agency thereof.
\end{acknowledgments}

\bibliography{apssamp}

\end{document}